\documentclass[a4paper]{spie}  
\usepackage[]{graphicx}
\usepackage{hyperref}
\usepackage{multirow}

\title{Monte Carlo simulations of Gamma-ray space telescopes:\\ a BoGEMMS multi-purpose application} 

\author{Valentina Fioretti\supit{*,a}, Andrea Bulgarelli\supit{a}, Marco Tavani\supit{b}, Martino Marisaldi\supit{a}, 
\\
Sabina Sabatini\supit{b}, Giuseppe Malaguti\supit{a}, Massimo Trifoglio\supit{a}, Fulvio Gianotti\supit{a}
\skiplinehalf
\supit{a}INAF/IASF Bologna, Via P. Gobetti 101, 40129, Bologna, Italy; \\
\supit{b}INAF/IAPS, via Fosso del Cavaliere 100, 00133 Roma, Italy
}

\authorinfo{* fioretti@iasfbo.inaf.it}
 
\begin{document} 
  \maketitle 

\begin{abstract}
After the development of a BoGEMMS (Bologna Geant4 Multi-Mission Simulator) template for the background study of X-ray telescopes, a new extension is built for the simulation of a Gamma-ray space mission (e.g. AGILE, Fermi), conceived to work as a common, multi-purpose framework for the present and future electron tracking gamma-ray space telescopes. The Gamma-ray extension involves the Geant4 mass model, the physics list and, more important, the production and treatment of the simulation output. From the user point of view, the simulation set-up follows a tree structure, with the main level being the selection of the simulation framework (the general, X-ray or gamma-ray application) and the secondary levels being the detailed configuration of the geometry and the output format. The BoGEMMS application to Gamma-ray missions has been used to evaluate the instrument performances of a new generation of Gamma-ray telescopes (e.g. Gamma-Light), and a full simulation of the AGILE mission is currently under construction, to scientifically validate and calibrate the simulator with real in-space data sets. A complete description of the BoGEMMS Gamma-ray framework is presented here, with an overview of the achieved results for the potential application to present and future experiments (e.g., GAMMA-400 and Gamma-Light). The evaluation of the photon conversion efficiency to beta particle pairs and the comparison to tabulated data allows the preliminary physical validation of the overall architecture. The Gamma-ray module application for the study of the Gamma-Light instrument performances is reported as reference test case. 

\end{abstract}



\section{INTRODUCTION}
\label{sec:intro}  
The Geant4 Monte Carlo toolkit \cite{g4_1, g4_2} is an open source, C++ based, particle transport code, initally developed by CERN for the simulation of high energy experiments involved at particle accelerators and then extended to ``lower" energy ranges, i.e. the X-ray and Gamma-ray domain. Geant4 is now a widely used particle transport code in the simulation of high energy instruments and radiation shielding optimisation of Astrophysics space missions. In terms of operative missions, it has been used for the simulation of the background induced radiation of X-ray telescopes like Chandra, XMM-Newton\cite{nartallo2001} and Suzaku\cite{2012SPIE.8443E..56O} or the scientific performances of the Fermi Gamma-ray mission\cite{2002APS..APRB17048M}. Many more are the Geant4-based simulations if we also count all the proposed projects (e.g. Simbol-X, IXO for the past, Athena, ASTRO-H, eRosita, GAMMA-400, and Gamma-Light for the future).
\\
The possibility to create a virtual model of an entire high energy space mission has fundamental benefits along the development process: from the instrument design to the calibration phase and the analysis of observational data.
Each time a new mission is proposed, or the mission design is dramatically changed, a brand-new Monte Carlo software project is usually built, including the code engineering of the satellite mass model, the physics processes set-up, the handling of the output (archiving and analysis). The major issues of such common approach are: (i) the simulation campaign is time and money consuming, (ii) a new validation of the physics and software architecture is required for each project, (iii) the simulation results are irreproducible because the software project is not maintened and shared after the mission operations/development end.
\\
The Bologna Geant4 Multi-Mission Simulator (BoGEMMS\cite{2012SPIE.8453E..35B}) has been conceived at the INAF/IASF Bologna as a modular and parameterized tool for the evaluation of the instrument performances (e.g., background induced radiation, detection efficiency, angular resolution) of X-ray and Gamma-ray space telescopes. The BoGEMMS software project includes both the Geant4-based simulator architecture and the filtering/analysis suite, with the aim of building a virtual model of the satellite and reproducing the real on-flight data taking. Its key feature is the possibility to interactively customize the tridimensional mass model, the physics processes, and the simulation output at run-time using a formatted configuration file\footnote{The set-up of the input particle spectra at run-time is achieved by means of the General Particle Source, an already built-in feature of the Geant4 toolkit.}, in order to potentially simulate any high energy mission using the same software architecture. It has been used\cite{fioretti_phd} for the study of the shielding optimization of the Simbol-X\cite{ff08} and the New Hard X-ray Mission\cite{gt09} (NHXM) telescopes, the evaluation of the soft proton funneling by the XMM-Newton\cite{2001A&A...365L..18S} and Simbol-X optics\cite{2008SPIE.7011E..86S}, and the study of the X-ray background radiation in Low Earth Orbit\cite{2012SPIE.8453E..31F}.
\\
Although the main requirement of the BoGEMMS simulator is the maximum flexibility of the geometry configuration without modifying the hardcoded libraries, some fields of applications require a detailed modeling of specific instruments and/or the analysis of a particular material property. An application-dependent simulation framework can be built in addition to the general, totally customizable, BoGEMMS environment. From the user point of view, the simulation set-up follows a tree structure, with the main level being the selection of the simulation framework (the general, X-ray or gamma-ray application). 
After the development of a BoGEMMS template for the background evaluation of focusing X-ray space telescopes, a new extension is built for the simulation of an electron tracking Gamma-ray space mission  (e.g., AGILE\cite{2009A&A...502..995T}, GAMMA-400\cite{2013AdSpR..51..297G}), conceived to work as a common, multi-purpose framework for the present and future gamma-ray space telescopes. Section \ref{sec:bogemms} describes in detail the Gamma-ray module design. Since its first application is the preliminary simulation of the Gamma-Light telescope\cite{2013NuPhS.239..193M}, we will refer to it as the test case for the present document. An overview of the achieved results for the potential application of present and future experiments is presented in Section \ref{sec:agile}.

\section{BOGEMMS ARCHITECTURE: the Gamma-ray module} \label{sec:bogemms}
The Gamma-ray regime for space-based telescopes ranges from hundreds of keV to tens of GeV. Above these energies, the amount of absorbing material that can be carried by the satellite payload is not enough to compensate the photon high penetration efficiency, and other on-ground detection techniques are applied (e.g. the Cherenkov telescopes). 
The BoGEMMS Gamma-ray module has been designed for the simulation of an electron tracking based telescope operating in the 1 MeV - 50 GeV energy range, composed by three key elements: the electron/positron tracker, the highly absorbing calorimeter and the anticoincidence (AC) shielding system. 
\\
In the tens of MeV, the Compton scattering is the dominant process: the incoming gamma-ray photon, interacting with a low Z material, releases part of its energy and momentum to an unbound electron. The tracker consists of several Silicon (Si) layers able to detect the position of the initial interaction and the energy and direction of the recoil electron. These measurements, coupled with the energy and direction of the scattered Gamma-ray photon collected by the high Z scintillation detector (the calorimeter) placed below the tracker, allow to indirectly measure the energy and direction of the primary photon. The use of electron tracking Compton telescopes is a very recent and still in progress technique, and it is the core of the proposed MEGA\cite{2005SPIE.5898...34B} and Gamma-Light telescopes. 
\\
Above 30-40 MeV, the pair production comes into play, with the photon annihilating into an electron/positron pair, the threshold for the interaction being 1.02 MeV ($\sim\rm 2m_{\rm e}$), with the energy of the photon taken up by the pair as rest mass and kinetic energy. In this case, a high Z convertion foil (e.g. Tungsten) is applied to the tracker Silicon planes. After the photon conversion, the sequence of Si planes tracks the path of the electron/positron pair and their secondaries, until they are absorbed by the calorimeter. Although the energy deposit in the calorimeter helps to better constrain the detection, the direction and energy mapping by the tracker can alone be enough to reconstruct the initial photon energy and direction. Both the AGILE and Fermi Gamma-ray telescopes use semiconductor-based pair trackers, and the same technology is planned for the GAMMA-400 mission. 
\\
The most challenging issue of Gamma-ray space telescopes is the minimization of the background induced radiation, because of the low signal-to-noise ratio. The causes are: (i) the intrinsic low photon rate; (ii) the cosmic ray dominance of the radiation environment; (iii) the background scales with the (extended) detection area. For this reason, tracker and calorimeter are surrounded by the AC active shielding. If a particle triggers both the AC and the Gamma-ray detectors, it is flagged as background and vetoed from the final event list.
\\  
The Gamma-ray extension not only involves the Geant4 mass model, but also the physics list and the production and treatment of the simulation output. Table \ref{tab:tree} reports the general schema of the configuration tree designed for the BoGEMMS Gamma-ray module.
At the time of writing, the BoGEMMS Gamma-ray release runs on both Geant4 9.1 and 9.6 versions.

\begin{table}[!h]
\centering
\begin{tabular}{c|c|c|c|c}
\multicolumn{5}{c}{\textsc{BoGEMMS Gamma-ray Module Tree}}\\
\hline
\hline
\multirow{26}{*}{Geometry}  & \multirow{12}{*}{Tracker}& \multirow{4}{*}{Strip} &  \multirow{1}{*}{On/Off} &\\
& & & \multirow{1}{*}{Replicated/Multiple} &  \\
& & & \multirow{1}{*}{Number} &  \\
& & & \multirow{1}{*}{Pitch} &  \\
\cline{3-5}
& & \multirow{8}{*}{Tray}& \multirow{1}{*}{Type} &  \\
& & & \multirow{1}{*}{Number} &  \\
& & & \multirow{1}{*}{Side} &  \\
& & & \multirow{1}{*}{Distance} &  \\
& & & \multirow{1}{*}{Passive material On/Off} &  \\
& & & \multirow{1}{*}{Converter On/Off} &  \\
\cline{4-5}
& & & \multirow{2}{*}{Layers} &  \multirow{1}{*}{Material}\\
& & &  &  \multirow{1}{*}{Thickness}\\
\cline{2-5}
& \multirow{5}{*}{Calorimeter} & \multirow{1}{*}{On/Off} & & \\
\cline{3-5}
&  & \multirow{3}{*}{Type} & \multirow{3}{*}{Pixel/Bar}& \multirow{1}{*}{Side}\\
&  &  & & \multirow{1}{*}{Height}\\
&  &  & & \multirow{1}{*}{Total side}\\
\cline{3-5}
&  & \multirow{1}{*}{Distance} & & \\
\cline{2-5}
& \multirow{4}{*}{Electronics} & \multirow{1}{*}{On/Off} & & \\
&  & \multirow{1}{*}{Type} & & \\
&  & \multirow{1}{*}{Distance} & & \\
&  & \multirow{1}{*}{Thickness} & & \\
\cline{2-5}
& \multirow{6}{*}{AC} & \multirow{1}{*}{On/Off} & & \\
\cline{3-5}
&  & \multirow{3}{*}{Type} & \multirow{1}{*}{Case A}& \\
&  &  & \multirow{1}{*}{Case B}& \\
&  &  & \multirow{1}{*}{Case C}& \\
\cline{3-5}
&  & \multirow{1}{*}{Distance} & \multirow{1}{*}{Top/Lateral/Bottom}& \\
\cline{3-5}
&  & \multirow{1}{*}{Thickness} & & \\
\hline
\multirow{2}{*}{Physics List} & \multirow{1}{*}{Type} & & & \\
\cline{2-5}
 & \multirow{1}{*}{Low Energy Electromagnetic} & \multirow{1}{*}{On/Off}& & \\
\hline
\multirow{7}{*}{Output} & \multirow{4}{*}{Type} & \multirow{1}{*}{XYZ}& & \\
 &  & \multirow{1}{*}{IN-OUT}& & \\
 &  & \multirow{1}{*}{ENERGY}& & \\
 &  & \multirow{1}{*}{STEP}& & \\
\cline{2-5}
 & \multirow{1}{*}{Volume selection} & & & \\
\cline{2-5}
 & \multirow{1}{*}{FITS rows number} & & & \\
\cline{2-5}
 & \multirow{1}{*}{FITS column selection} & & & \\
\hline
\multirow{7}{*}{Visualization} & \multirow{3}{*}{Tracker} & \multirow{1}{*}{Coloured}& & \\
 &  & \multirow{1}{*}{Black/White}& & \\
 &  & \multirow{1}{*}{Only Si planes}& & \\
\cline{2-5}
 & \multirow{4}{*}{On/Off} & \multirow{1}{*}{Calorimeter}& & \\
 &  & \multirow{1}{*}{Electronics}& & \\
 &  & \multirow{1}{*}{AC}& & \\
 &  & \multirow{1}{*}{AC top panel}& & \\
\hline
\end{tabular}
\caption{\label{tab:tree}The tree structure of the configuration file for the simulation of Gamma-ray space telescopes using BoGEMMS.}
\end{table}

\subsection{Geometry}\label{sec:geo}
In terms of geometry, the user must first select the appropriate keyword for the Gamma-ray module simulation. Three main blocks, each acting as an independent detection system, constitute the geometry: the tracker (Section \ref{sec:tracker}), the calorimeter (Section \ref{sec:cal}), and the AC shielding system (Section \ref{sec:ac}). The electronics cards that laterally surround the tracker are also added. The code is designed around the tracker, which is fixed: the calorimeter and the anticoincidence system can be added or removed from the environment at run time. 
\\
The BoGEMMS Gamma-ray extension has been applied to evaluate the scientific performances of the Gamma-Light mission in range of the 2012 ESA call for a small mission opportunity\cite{esa_proposal}. For this reason, the Gamma-Light simulation has been chosen as test case for the present document, and all the Geant4 mass model images presented here refer to the Gamma-Light design. Supported by a joint Italian-European high energy Astrophysics community, the telescope goal is to observe the 10--100 MeV energy range with unprecedented sensitivity ($<2-3\times10^{-6}$ cm$^{-2}$ s$^{-1}$ MeV$^{-1}$ at 10 MeV, for $\Delta\rm T=1$ Ms) and angular resolution ($1^{\circ}-2^{\circ}$ at 10 MeV), about an order of magnitude better than the COMPTEL instrument on board CGRO\cite{1994ApJS...92..351G}.

\subsubsection{Tracker}\label{sec:tracker}
   \begin{figure}
   \begin{center}
   \begin{tabular}{c}
   \includegraphics[width=0.95\textwidth]{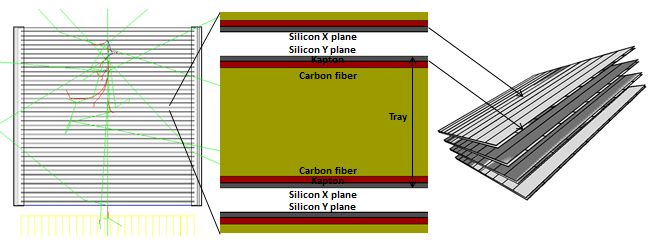}
   \end{tabular}
   \end{center}
   \caption[tracker] 
   { \label{fig:tracker} \textit{Left panel}: Lateral view of the photons (in green), electrons (in red) and positrons (in blue) generated by a 10 MeV monochromatic beam (10 primary photons) parallel to the tracker axis. Only the Si planes are visualized in grey, the lateral bars represent the electronic cards surrounding the tracker and the calorimeter is shown in yellow at the bottom. \textit{Central panel}: The zoomed view of a tray (the white regions are the few millimeters separation between the trays). \textit{Grey}: Si layer; \textit{black}: Kapton layer; \textit{red}: Carbon fiber layer; \textit{yellow}: Aluminum honeycomb layer. \textit{Right panel}: the strips composing the Si X/Y views.}
   \end{figure} 
The tracker is currently designed as one AGILE-like\cite{2010NIMPA.614..213B} single column, composed by several trays according to the user-defined configuration. A tray is defined here as the tracker unity block, and it includes the two Si planes, the X and Y layers for the hit position mapping, and the passive structural material. Figure \ref{fig:tracker} (left panel) shows the Si planes in grey and the wireframed calorimeter in yellow, placed at the bottom. A 10 MeV monochromatic beam, parallel to the tracker axis, is simulated (the green, red and blue lines refer to the photons, electrons and positrons respectively). The tray is zoomed in the right panel, also showing this time the passive layers, with the Si, Kapton, Carbon fiber and Aluminum (Al) honeycomb layers visualized in grey, black, red and yellow respectively. Only the Si planes are Geant4 sensitive volumes. 
\\
The general tray configuration is based on the AGILE/GRID tracker design, but the converter foil, composed in the AGILE case by Tungsten, can be removed if a low energy Gamma-ray telescope is simulated, as the Gamma-Light case. The X and Y Si views are divided into microstrips that define the instrument tracking resolution (see Figure\ref{fig:tracker}, right panel). As example, the AGILE views are composed by 3072 strips, 0.121 mm side, counting both read-out and floating strips: a total number of 73728 volumes for the Si planes is required, which can be quite CPU time consuming. For this reason, the strip can be created using the replicated Geant4 volumes to reduce the impact on the simulation performances. The user can set size, thickness, materials, and number of the tracker trays, as well as the number of Silicon strips dividing the X/Y planes.
\\
The electronic cards at the lateral sides are basically Al boxes with a Si layer placed inside, to take into account the disturbance effect induced by the surrounding passive material to the particle tracks. Only the general configuration (dimension and distance from the tracker) can be configured (see Figure \ref{fig:cal}, right panel).


\subsubsection{Calorimeter}\label{sec:cal}
\begin{figure}
   \begin{center}
   \begin{tabular}{c}
   \includegraphics[height=6cm]{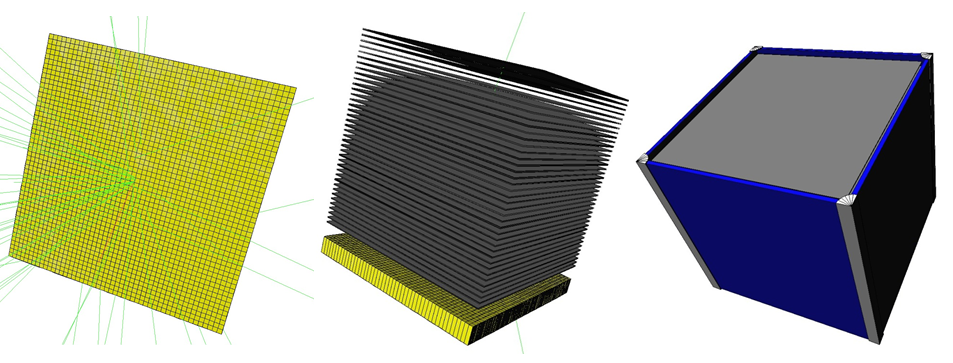}
   \end{tabular}
   \end{center}
   \caption[cal] 
   { \label{fig:cal}\textit{Left panel}: the CsI(Tl) calorimeter divided into 1 cm side pixels illuminated by a 1 GeV test beam. \textit{Central panel}: view of the tracker Si X/Y planes (in grey) and, at the bottom, the pixellated calorimeter (in yellow). \textit{Right panel}: the electronic cards (in blue) placed at the tracker lateral sides plus the supporting Al columns.}
   \end{figure}

The calorimeter is an inorganic scintillation detector, in which light is produced by ionizing radiation (e.g., Gamma-rays and beta particles, but also cosmic rays) and collected by a phototube where the light is converted into an electrical signal. At present, Thallium activated Cesium Iodide (CsI(Tl)) is the fixed calorimeter composition, but the possibility for the material selection at run-time will be added in the future. The calorimeter is position sensitive, and two configurations are foreseen: pixel subdivision (e.g., Gamma-Light) and perpendicular bars placed into two planes (e.g., AGILE), acting as the X and Y coordinates of the hit. The pixellated configuration (1 cm pixel side) illuminated by a 1 GeV test photon beam, visible in green, is shown in Figure \ref{fig:cal}, left panel. The calorimeter is totally parameterized, so that the user can choose the pixel/bar side, thickness, and distance from the tracker.
\subsubsection{Anticoincidence system}\label{sec:ac}
   \begin{figure}
   \begin{center}
   \begin{tabular}{c}
   \includegraphics[height=10cm]{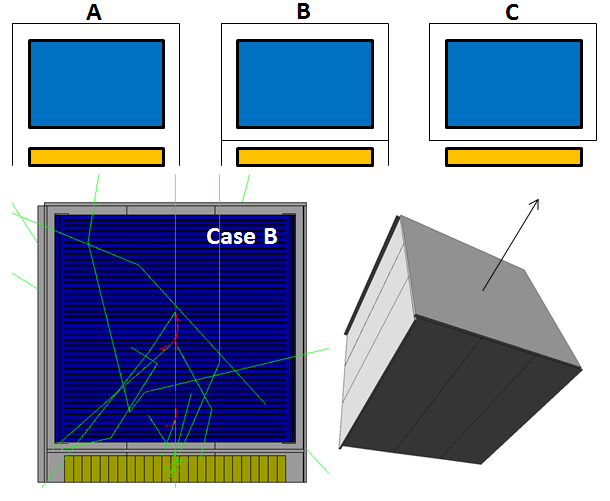}
   \end{tabular}
   \end{center}
   \caption[ac] 
   { \label{fig:ac} The AC shielding system design conceived for the Gamma-ray extension. \textit{Top panels}: the three configurations labelled as case A, B, and C. The blue and yellow boxes refer to the tracker and calorimeter respectively. \textit{Bottom panels}: the lateral cut view of the whole telescope (on the left) illuminated by a 10 MeV test beam and the AC grey structure (right panel), with the arrow indicating the field of view direction. The lateral subpanels are also visible.}
   \end{figure} 
The AC system is based on the AGILE shielding design\cite{2006NIMPA.556..228P} and it is composed by Bicron BC-404 plastic scintillation panels: three panels at each lateral side, one panel for the top. The design of the AC system greatly impact on the instrument performances. Not only its background rejection efficiency is one of the key technical requirements and any particle leakage must be avoided, but it can also cause auto-veto because of particle backscattering from the calorimeter or secondary interactions from electrons produced in the tracker.
For this reason, in addition to the general flexibility in terms of thickness and distances, three different configurations are foreseen for the AC system:
\begin{itemize}
\item[A:] the lateral panels are extended at the bottom to only cover the calorimeter (e.g. AGILE);
\item[B:] in addition to the case A design, an horizontal AC panel is placed between the tracker and the calorimeter (e.g. Gamma-Light);
\item[C:] the calorimeter is not shielded by the AC while the tracker is totally covered by the panels.
\end{itemize}
The three cases are schematized in Figure \ref{fig:ac}, top panels. The case B lateral cut view and the total AC structure, in grey, are also shown at the bottom. With the use of a simple flag, the impact of the background induced radiation on the performances of a Gamma-ray telescope can be easily computed as a function of the AC design.

\subsection{Physics lists}\label{sec:phys}
In terms of physics processes, a keyword allows to easily select between both Geant4 reference physics lists\footnote{\url{http://geant4.cern.ch/support/proc_mod_catalog/physics_lists/referencePL.shtml}} (e.g., QGSP$\_$BERT$\_$EMV for the use of quark gluon string model and Bertini cascades for hadronic interactions combined with optimized electromagnetic processes) and composite lists (e.g. the Fermi-based Space Sciences Physics List provided by the SLAC institute\footnote{\url{http://www.slac.stanford.edu/comp/physics/geant4/slac_physics_lists/space/glast_physics_list.html}} or the Gamma-ray telescope advanced example included in the Geant4 release). Not all the physics lists are always available, depending on the Geant4 version that has being used. In the case of Geant4 9.1, a flag can be used for the activation of the low energy electromagnetic packet. The results presented here are obtained using the Fermi-based SLAC physics list.

\subsection{Output data format}\label{sec:output}
How the simulation results (e.g. the energy deposits on the sensitive volume, particle energy) are selected and formatted for data storage is probably the most important feature of a Monte Carlo simulator, because it affects the CPU processing time and the filtering/analysis pipeline. A new BoGEMMS simulation output is created for the simulation of Gamma-ray space telescopes, is addition to the ENERGY, IN-OUT, and STEP types\cite{2012SPIE.8453E..35B}.
\\
The new output file, in FITS format, which collects the hits of the activated sensitive volumes (tracker, calorimeter, and AC), is defined as XYZ because specifically designed for position sensitive energy deposits.
The particle history is recorded from the volume point of view: until its total absorption, the particle properties are saved at the entrance and exit (or last interaction step) of the volume, plus the total energy deposit in the volume. If a secondary is produced within the sensitive volume, its energy deposit is also recorded along its path. A new feature in the sensitive detector class has been added for the analysis of tracks in Compton/pair production telescopes. When a secondary is produced, if it is generated by a Compton scattering or a Gamma-ray conversion a flag is added according to the process. In addition, both primary and secondary photons that Compton scatter in the active tracker layers are flagged in the output file. 
\\
In the configuration file is possible to select which volumes writing in output. Consistently with the analysis pipelines applied in operative Gamma-ray missions, the set of activated tracker, calorimeter, and AC volumes (and related hits) are collected and cross-correlated by the filtering process to remove the vetoed events and apply the trigger patterns.
\\
The following parameters are written:
\begin{itemize}
\item EVT$\_$ID = event number (starting from 0)
\item TRK$\_$ID = track number (starting from 1)
\item VOLUME$\_$ID = volume number
\item Given N the number of trays, the tray ID is $\rm N\times1000$ (starting from the tracker bottom).
\item Given Nlayer the layer number (from the tray bottom), the layer ID, in each tray, is the tray ID + (N$_{\rm layer}$ – 1).
\item VOLUME$\_$NAME = volume name string
\item E$\_$DEP = energy deposit (in keV)
\item X$\_$ENT, Y$\_$ENT, Z$\_$ENT = position, in mm, of the particle entrance (or generation) point in the sensitive volume
\item X$\_$EXIT, Y$\_$EXIT, Z$\_$EXIT = position, in mm, of the particle exit (or final) point in the sensitive volume
\item E$\_$KIN$\_$ENT = particle kinetic energy, in keV, at the entrance (or generation) point
\item E$\_$KIN$\_$EXIT = particle kinetic energy, in keV, at the exit (or final) point
\item MDX$\_$ENT, MDY$\_$ENT, MDZ$\_$ENT = particle direction cosines at the entrance (or generation) point
\item MDX$\_$EXIT, MDY$\_$EXIT, MDZ$\_$EXIT = particle direction cosines at the exit (or final) point
\item GTIME$\_$ENT = simulation global time at the particle entrance (or generation) point
\item GTIME$\_$EXIT = simulation global time at the particle exit (or final) point
\item PARTICLE$\_$ID = particle identification number according to the Particle Data Group
\begin{itemize}
\item[-] Gamma = 22
\item[-] Electron = 11
\item[-] Positron = -11
\item[-] Proton = 2212
\item[-] Neutron = 2112
\end{itemize}
\item PARTICLE$\_$NAME = particle name string
\item PROCESS$\_$FLAG = flag for Compton/pair production identification
\end{itemize}

\section{APPLICATION TO PRESENT AND FUTURE GAMMA-RAY MISSIONS}\label{sec:agile} 
The BoGEMMS software project includes the simulation analysis for the evaluation of the instrument scientific performances, a process which is highly mission-dependent. However, the event cleaning algorithms, e.g. the hit selection and the signal digitization, are the same for all the Gamma-ray telescopes. For this reason, we built a parameterized, IDL and Python based, pipeline where the user selects the mission type, and the mission-dependent parameters (e.g. the tracker geometry, the AC minimum threshold) are loaded accordingly. The preliminary simulation validation obtained for the AGILE telescope is presented in Section \ref{sec:agilesim} and Section \ref{sec:det_eff} reports some results from our test case, the Gamma-Light mission, using the BoGEMMS analysis pipeline.

\subsection{Filtering pipeline}\label{sec:filt}
The filtering and analysis of the simulation output follows several steps, according to the required level of data cleaning.
First of all, the energy deposits are selected from the Si planes, AC panels and calorimeter pixels/bars (a unique identification number is associated to each volume). This is the so-called raw level. The data level 0 of the filtering pipeline consists in summing all the energy deposits in the same volume, and sorting the hits according to the tray number for each same simulated event. From the simulation output, we obtain the position of the entrance and exit/absorption points: this information is substituted by the strip central position coordinates, as in the real case.
\\
The events digitization, i.e. the application of the strip capacitive coupling and detection threshold, takes into account the effect of the electronic system on the detected energy deposit and results in the data level 0.5. The capacitive coupling is the energy transfer between the tracker strips due to the electron cloud generated by the hit. All the strips are built in the Geant4 mass model, so that different read-out patterns can be applied. For the BoGEMMS Gamma-ray extension analysis, we choose the analog read-out of the AGILE telescope, which ensures an outstanding point spread function less than $1^{\circ}$ at 1 GeV\cite{2013A&A...558A..37C}. In the analog read-out, only one Si strip out of two is read by the dedicated TA1 chips. The energy deposit in the floating strips is inferred by means of the capacitive coupling that adds a fraction of this energy to the contiguous strips\cite{2002NIMPA.486..610L}. The final energy deposit in each read-out strip results from the weighted sum of all its neighbours. The fraction of energy transferred by the capacitive coupling depends on the tracker design, and must be measured using laboratory test beams. For the Gamma-Light simulation, we used the strip charge redistribution of the AGILE telescope\cite{2002NIMPA.490..146B}. After the simulation of the capacitive coupling, we apply a minimum detection threshold of 0.25 Minimum Ionising Particles ($\sim27$ keV). The 0.5 data level is saved in both FITS and ASCII format. The ASCII files are then given in input to the IDL-based Kalman filter\cite{2006NIMPA.568..692G} for the computation of the instrument point spread function.
\\
From data level 1, a sequence of track pattern recognition filters, with increasing complexity, is applied:
\begin{itemize}
\item[1:  ] only the energy deposits present in both the X and Y layers of the same tracking plane are saved;
\item[1.5:] only the tracks hitting three X/Y planes are saved;
\item[2:  ] only the tracks hitting three X/Y planes out of four contiguous planes are saved.  
\end{itemize}
At the end of the filtering pipeline, six data levels are saved in FITS and ASCII format files for the analysis and/or visualization later steps.

\subsection{The AGILE simulation}\label{sec:agilesim}
\begin{figure}
   \begin{center}
   \begin{tabular}{c}
   \includegraphics[width=0.48\textwidth]{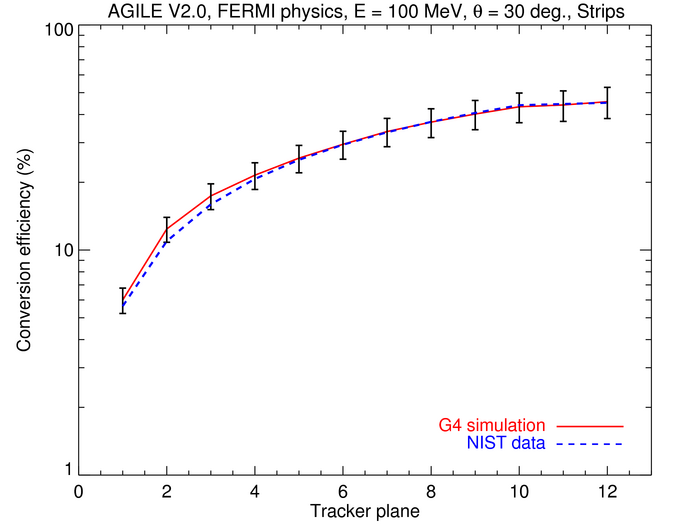}
   \includegraphics[width=0.48\textwidth]{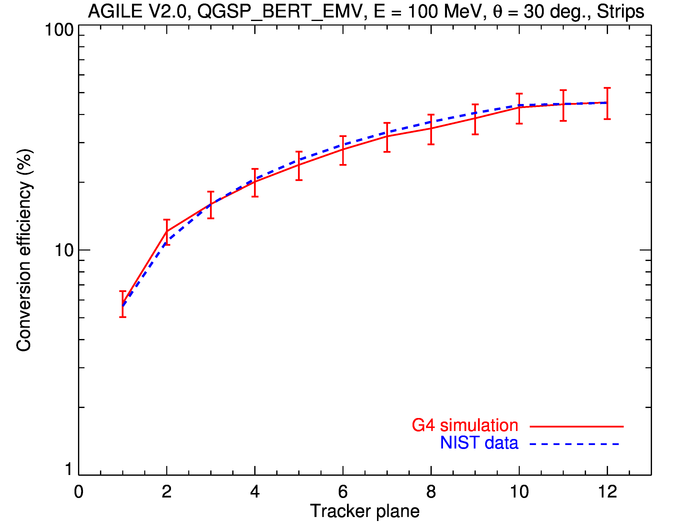}
   \end{tabular}
   \end{center}
   \caption[agile] 
   { \label{fig:agile} The efficiency, in \% of the input photon number, of Gamma-ray photon conversions to electron/positron pairs as a function of the number of crossed tracker planes. \textit{Red continous line}: BoGEMMS Geant4 simulation; \textit{blue dashed line}: the computed efficiency using the attenuation coefficients of the NIST agency. \textit{Left panel}: the Fermi-like SLAC physics list is used with Geant4 9.1. \textit{Right panel}: the QGSP$\_$BERT$\_$EMV reference Geant4 physics list is used with Geant4 9.6.}
   \end{figure}

In order to use BoGEMMS for the simulation of future Gamma-ray space experiments (e.g. the evaluation of the GAMMA-400 point spread function) the simulation must be physically validated by comparing the results with both on-ground and on-flight experimental data. Since the whole telescope geometry and the filtering pipeline is based on the AGILE mission, the ability to reproduce the AGILE instrument perfomances, i.e. the point spread function and the effective area but also the background level, would physically prove the correct design of the Gamma-ray simulation framework. This comparison would define the BoGEMMS software project as the reference simulation tool for future electron tracking Gamma-ray telescopes. In addition, once the simulation is calibrated to the real on-flight data sets, a detailed analysis of the effect of the reference Geant4 physics lists on the simulation accuracy is foreseen.
\\
The full AGILE telescope simulation must include the use of the operative AGILE filter\cite{agile_filter} and analysis pipeline after the preliminary filtering of Section \ref{sec:filt}. This activity is currently in progress. A first and easy step in the Gamma-ray module validation is to analytically compute a physical property on the base of tabulated parameters, and compare it to the simulation results. Using the X-ray and Gamma-ray attenuation coefficients provided by the National Institute of Standards and Technology agency (NIST\footnote{\url{http://www.nist.gov}}), we compute the efficiency of producing an electron/positron pair from incident 100 MeV photons, simulated for an inclination angle, in polar coordinates, of $\theta=30^{\circ}$, $\phi=225^{\circ}$, as a function of the crossed number of Si planes.
\\
The resulting comparison with the AGILE tracker simulation, in \% of the input photons, is reported in Figure \ref{fig:agile}, using the Fermi-like SLAC (Geant4 9.1) and QGSP$\_$BERT$\_$EMV reference Geant4 physics list (Geant4 9.6) in the left and right panels respectively.
Since in the simulation we take the first hit on a Si layer, we can only compute the cumulative efficiency, i.e. the efficiency of producing a conversion at a plane lower than the triggered layer (the plane number starts from the sky side of the tracker). For example, a conversion efficiency of 10\% at the plane 2 means that the 10\% of the incoming photons interact before the plane 2.
\\
Within the simulation intrinsic Poisson fluctuations, the BoGEMMS simulation is consistent with the computed values for both the applied physics list, althought the reference list seems to better reproduce the interaction for a low number of planes. A maximum efficiency of $\sim40\%$ is reached.

\subsection{Gamma-Light efficiency and background rejection}\label{sec:det_eff}
   \begin{figure}
   \begin{center}
   \begin{tabular}{c}
   \includegraphics[width=0.48\textwidth]{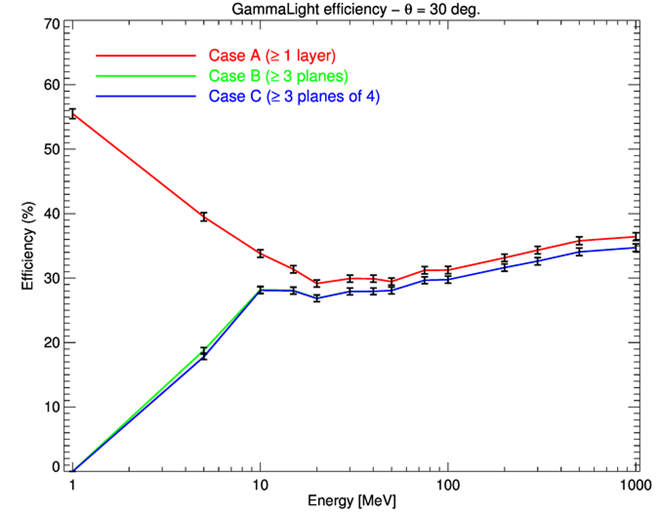}
   \includegraphics[width=0.48\textwidth]{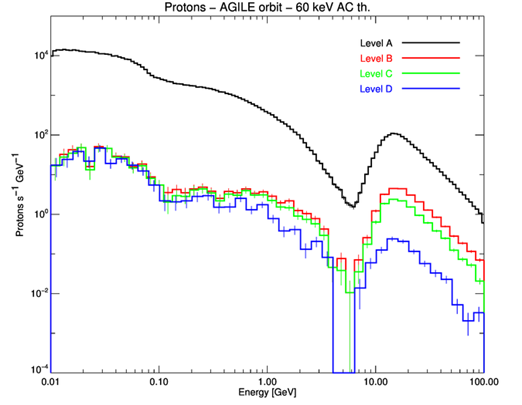}
   \end{tabular}
   \end{center}
   \caption[eff] 
   { \label{fig:eff} 
\textit{Left panel}: The detection efficiency (in \% of the simulated events) for the case A (red line), B (green line) and C (blue line) pattern recognition type, for an incoming photon angle of $30^{\circ}$. \textit{Right panel}: The proton composition of the GCR and Earth albedo flux, integrated for the total exposed spacecraft area, causing a detection on the tracker for different AC triggering patterns.}
   \end{figure}
The tracker efficiency in detecting the input Gamma-rays is computed for the Gamma-Light telescope. The tracker recontructed events are then cross-checked and removed by the AC system for different trigger algorithms, and a preliminary efficiency in background proton rejection is computed.
\\
The detection efficiency is generally defined as $\rm Eff=N_{\rm det}/\rm N_{\rm in}$, where $\rm N_{\rm det}$ and $\rm N_{\rm in}$ are the number of detected events and input particles respectively. Since Gamma-rays can not be directly detected, but must be observed by means of the secondary particles produced by Compton scattering and/or pair produc tion, the detection efficiency depends not only on the instrument sensitivity but also on the accuracy of the recognition pattern.  
\\
The efficiency is evaluated for the following energies: 1, 5, 10, 20, 30, 40, 50, 75, 100, 300, 500 MeV and 1 GeV. The photons are simulated as monochromatic beams for an inclination angle, in polar coordinates, of $\theta=30^{\circ}$, $\phi=225^{\circ}$. if $\theta=0^{\circ}$, the photons are directed along the tracker axis.
\\
Three different patterns are applied in the evaluation of the Gamma-Light efficiency. The incoming photon is counted as detected ($\rm N_{\rm det}$) if:
\begin{itemize}
\item[A:] it generates at least one hit (energy deposit $>0$) in a Si layer (X and/or Y);
\item[B:] it is detected by at least three planes (X and Y Si layers) in the whole tracker;
\item[C:] it is detected by at least three planes (X and Y Si layers) of four contiguous planes.
\end{itemize}
Each efficiency case study represents a subset of the previous case.
These patterns, especially cases A and B, are raw approximation of the complex data analysis, given the simplified geometry and the simulation level. Although preliminary, these results are a first, fundamental indication of the potential scientific performances of Gamma-Light. The resulting efficiency, in \% of the input photons, is presented in Figure \ref{fig:eff} (left panel). As expected, the red curve follows the decrease of the Compton scattering probability, while the occurance of three planes triggers dramatically increases below 10 MeV and reaches a plateau at higher energies, as the evolution of the pair production interaction dominance. In the 10-1000 MeV energy range, the Gamma-Light tracker detects about the 30-40\% of the input Gamma-ray photons.
\\
A Low Earth Orbit (LEO) with low inclination (AGILE-like) has been proposed for the Gamma-Light launch. Assuming the AGILE orbit, the background proton input flux is given by the sum of the Galactic Cosmic rays (GCRs) and the Earth albedo proton flux\cite{2012SPIE.8453E..31F}. In particular, the GCR flux is analitically computed\cite{2004ApJ...614.1113M} for a solar minimum of $\Phi=500$ MV, following a conservative approach. A $70\times70\times150$ cm box with an Aluminum equivalent mass of 200 kg, is placed below the instruments to take into account the effect of secondary proton interactions with the satellite payload. The flux, integrated along the total spacecraft area including the payload, is plotted in Figure \ref{fig:eff} (right panel, black line). The protons interact with all the material surrounding the tracker, and their secondaries cause hits on the tracker planes that can be assigned to astrophysical events. For this reason, the hits on the AC panels (a minimum detection threshold of 60 keV is applied) are also analyzed and used for background rejection. The following triggering patterns are applied:
\begin{itemize}
\item[A:] the AC shielding system is turned off;
\item[B:] same as level A, but the incoming protons must activate three of four contiguous planes in the tracker;
\item[C:] the protons of level B that do not trigger the AC top panel;
\item[D:] the protons of level B that do not activate any AC panel.
\end{itemize}
The only use of the simple track recognition of level B removes about a factor 100 and 10 of the incoming proton flux below and above 1 GeV respectively. The AC panels mostly reject the high energy protons ($>10$ GeV). Since the payload passive material is placed below the instruments, the AC lateral panels trigger is mainly responsible of the proton flux reduction at high energies.


\section{Summary and future plans} 
A new BoGEMMS modular extension is being developed for the simulation of electron tracking Gamma-ray telescopes, that allows the user to configure the geometry, the physics lists and the output format at run-time. The Gamma-ray framework is specifically designed to evaluate, using the same software project, the instrument performances of multiple mission designs. 
\\
The BoGEMMS extension has been used for the simulation of the Gamma-Light telescope: the detailed description of the mass model and some simulation results are reported as reference test case of the Gamma-ray module.
A complete simulation of the AGILE telescope, including its interface with the real mission filter and scientific analysis pipeline, is currently under construction. The simulation of the AGILE on-ground and on-flight instrument performances is planned, to physically validate the BoGEMMS Gamma-ray extension and apply the same architecture for the simulation of next generation Gamma-ray telescopes. The preliminary validation using tabulated data is successfull, and we are able to reproduce the expected photon conversion efficiency. Once physically validated, the use of the BoGEMMS software project for the production of the calibration matrices of the AGILE telescope is planned. 
\\
The expertise achieved along these years in the study of the high energy space radiation environment, coupled with the ability to reproduce the experimental performances of the AGILE telescope, would make BoGEMMS the perfect tool for the simulation of the scientific performances of future X-ray and Gamma-ray missions. In the high energy regime, the ROSCOSMOS GAMMA-400 space observatory launch is planned in 2018: designed to detect both Gamma-rays, electron/positrons, and nuclei, its istruments include scintillation AC detectors, a W/Si converter/tracker system, and BGO crystals calorimeters. In the X-ray energy range, the ATHENA\cite{athena_esa_proposal} telescope has been recently proposed as the next ESA X-ray observatory, with launch in 2028. Operating below 15 KeV, it foresees a single mirror module and two focal plane instruments, a Wide Field Imager (WFI) and an X-ray Integral Field Unit (X-IFU) with high spectral resolution.


\bibliography{fioretti_spie_2014}   
\bibliographystyle{spiebib}   

\end{document}